\documentclass{article}

\PassOptionsToPackage{numbers,sort&compress}{natbib}
\usepackage[preprint]{neurips_2026}

\usepackage[utf8]{inputenc} 
\usepackage[T1]{fontenc}    
\usepackage{hyperref}       
\usepackage{url}            
\usepackage{booktabs}       
\usepackage{amsfonts}       
\usepackage{amsmath}
\usepackage{nicefrac}       
\usepackage{microtype}      
\usepackage{xcolor}         
\usepackage{tcolorbox}
\usepackage{wrapfig}
\usepackage{multirow}
\usepackage{graphicx}
\usepackage{subcaption}
\usepackage[table]{xcolor}
\definecolor{permitblue}{RGB}{235,243,250}
\usepackage{amsthm}
\newtheorem{observation}{Observation}
\usepackage{algorithm}
\usepackage{algpseudocode}
\usepackage{tikz}
\usetikzlibrary{positioning,arrows.meta}
\usepackage{tabularx}
\usepackage{array}
\usepackage{makecell}
\usepackage{colortbl}
\usepackage{caption}
\usepackage{soul}

\newcommand{\sens}[1]{\textcolor{red!80!black}{#1}}

\sethlcolor{green!20}
\newcommand{\allowinfo}[1]{\hl{#1}}

\title{\textsc{Permit:} Permission-Aware Representation Intervention for Controlled Generation in Large Language Models}

\author{Pengcheng Sun$^{1}$, Lan Zhang$^{1,2}$\thanks{Corresponding author.}\texttt{ }, Zhaopeng Zhang$^{1}$, Jiewei Lai$^{1}$, Chen Tang$^{1}$\\
$^{1}$University of Science and Technology of China,\\
$^{2}$Institute of Artificial Intelligence, Hefei Comprehensive National Science Center}

\begin{document}

\maketitle

\begin{abstract}
Large language models (LLMs) are increasingly deployed in enterprise settings where they handle sensitive documents and user context, raising acute concerns over security and controllability. Conventional access control regulates \emph{whether} information is accessible to the model, yet leaves \emph{how} the model uses that information at generation time largely unconstrained: once sensitive content enters the context, outputs may still drift beyond a user's authorized scope. We present \textsc{Permit}, a novel permission-aware representation intervention framework that closes this gap by enforcing fine-grained control directly on the model's hidden states. 
Through exploratory analysis, we find that permission conditions induce hidden-state shifts that are (i) separable across permissions and (ii) concentrated in a small set of dominant directions.
\textsc{Permit} exploits this geometry in two stages: it first identifies a \emph{permission-sensitive subspace} from activation differences across permission conditions, and then performs lightweight interventions within this subspace to steer generation, with two concrete instantiations (\emph{offset-based} and \emph{gated}). 
Both operate atop a frozen backbone with only a handful of permission-specific parameters, achieving precise control with minimal overhead. Experimental results demonstrate that \textsc{Permit} performs better than the state-of-the-art method across multiple permission settings while driving information leakage to near zero, achieving over \(18\%\) F1-score improvement with \(>98\%\) fewer trainable parameters.


\end{abstract}

\section{Introduction}

Large language models (LLMs) have demonstrated remarkable capabilities across a wide range of tasks~\citep{DBLP:journals/tist/ChangWWWYZCYWWYZCYYX24,DBLP:conf/aaai/Chai26}, fueling growing interest in deploying them within enterprise settings~\citep{DBLP:conf/sigmod/CaoFX0CY25, DBLP:conf/acl/RajendranDSC25, DBLP:conf/emnlp/VishwakarmaAPDC25} for applications such as knowledge query and data analysis. Yet processing sensitive information (e.g., internal documents, chat histories, proprietary records) introduces security risks that have become a major barrier to broader deployment~\citep{DBLP:conf/naacl/RamrakhiyaniMPAMSS25}. A foundational requirement is \emph{access control}: even when given the same input, users with different permissions must receive different, authorization-compliant outputs~\citep{DBLP:journals/pvldb/BodensohnBVSB25}. Real-world permission policies are typically fine-grained and hierarchical, and equipping LLMs to faithfully respect such structured policies remains a core open challenge~\citep{SudoLM,DOMBA,lazier2025aclora,DBLP:journals/corr/abs-2504-09593}.

Existing work falls broadly into two paradigms. \emph{Alignment-based} methods rely on supervised fine-tuning~\citep{Sodullm} or RLHF~\citep{SudoLM,RLHFnips} to instill globally safe behavior and refuse harmful requests~\citep{chen-etal-2024-combating}; while effective at coarse, binary safety boundaries, they struggle to faithfully model multi-role, hierarchical permissions~\citep{RBAC}. \emph{Rule-based} methods instead bind explicit policies to data, roles, or model components via policy-specific fine-tuning~\citep{DOMBA,DBLP:conf/ijcnlp/AlmheiriKSTVKK25}, parameter encryption~\citep{SECNEURON}, or per-domain LoRA adapters~\citep{lazier2025aclora,PermLLM,AdapterSwap}. These approaches more closely mirror traditional access-control systems, but rely on static partitioning that scales poorly as the number of roles and permission levels grows.

More fundamentally, both paradigms share a common blind spot. Once sensitive content enters the model's usable context—via retrieval-augmented generation, user input, or external tools—the model can still produce content beyond the user's authorized scope~\citep{ARBITER}. This exposes a key disconnect between whether information is made accessible to the model and how the model uses that information during generation. 
Existing methods regulate the former but offer little leverage over the latter.
Therefore, a central question is: \emph{how can we enforce fine-grained control over information use at generation time, across different permission levels, without sacrificing utility?}

Answering this question requires confronting two fundamental difficulties. \textbf{(C1) Entanglement.} Complex authorized and unauthorized content co-occur in the context and become entangled in the model's internal representations, making selective suppression hard without degrading fluency or factuality. \textbf{(C2) Security–utility tension.} Overly conservative control causes over-refusal, while insufficient control leaks information. Striking a fine-grained balance across many permission levels rather than a single global safety setting remains a challenging problem.


\paragraph{This work.}
We address this disconnect by enforcing permissions at the \emph{generation stage}, regulating how information is used rather than merely whether it is accessible. Instead of treating permissions as external rules, we seek internal representations that encode permission differences and use them to directly steer generation. This view is motivated by recent advances in activation engineering~\citep{liu2026perfit,DBLP:conf/iclr/Wang0025,DBLP:conf/iclr/StolfoBYHN25}, which show that targeted interventions on hidden states can reliably steer LLM behavior without modifying the backbone. Through exploratory analyses, we uncover two complementary phenomena: \textbf{(1) Activation-space separability:} hidden states under different permissions form well-separated clusters whose shifts reflect the hierarchical structure of the underlying policies; and \textbf{(2) Low-rank concentration:} these permission-induced shifts concentrate along a small number of dominant directions and are well approximated by a low-dimensional subspace. 

Building on these findings, we propose \textsc{Permit}, a permission-aware representation intervention framework for fine-grained controlled generation. 
\textsc{Permit} proceeds in two stages.
Specifically, it first learns a \emph{permission-sensitive subspace} from activation differences across permission conditions, capturing the principal directions of permission-related variation (addressing \textbf{C1}). Then it performs interventions within it via a unified subspace intervention framework with two instantiations: an \emph{offset-based} intervention re-aligns the desired representations through an affine transformation for control under strict permission boundaries, while a \emph{gated} intervention applies dimension-wise scaling for fine modulation along the security–utility frontier (addressing \textbf{C2}). Both keep the backbone LLM frozen and introduce only a small number of permission-specific parameters.

Overall, our \textbf{main contributions} are as follows:
(1) \emph{Perspective.} We reframe access control in LLMs as representation-space intervention at the generation stage, enabling plug-and-play enforcement and complementing existing data-level approaches.
(2) \emph{Method.} We show that permission-induced representation shifts are both separable and approximately low-rank, and propose a novel permission-sensitive subspace intervention framework for fine-grained access control.
(3) \emph{Results.} 
Extensive experiments show that \textsc{Permit} achieves a substantially better security--utility trade-off than baselines: it drives information leakage to near zero (\(0.0\%-2.5\%\)) while improving F1-score by over \(18\%\), with \(>98\%\) fewer trainable parameters.

\section{Preliminary}

\subsection{Problem Statement}

Let $\mathcal{R}=\{r_k\}_{k=1}^{N}$ denote a set of permission states, where each $r_k$ specifies structured access attributes such as user role, domain, and response policy.
For each permission state $r_k$, let
$\mathcal{D}_{k}=\{(q_i^{k}, c_i^{k}, y_i^{k})\}_{i=1}^{n_k}$
be the corresponding permission-conditioned dataset, where $q_i^{k}$ is a query, $c_i^{k}$ is the available context, and $y_i^{k}$ is the permission-compliant response. Here, $n_k$ denotes the number of queries for $r_k$.
Given a base LLM $M_0$ with parameters $\Theta_0$, the model produces a generic output
\(
\hat{y}_i^{\,0} = M_0(q_i^{k}, c_i^{k}),
\)
which does not explicitly enforce permission-aware semantic control.
As a result, the model may reveal unauthorized information by using, combining, or inferring sensitive content from the context or its parametric knowledge.

\textbf{Our goal} is to learn a lightweight permission-conditioned intervention module $\Delta_k$ for each permission state $r_k$.
The resulting controlled model generates
\(
\hat{y}_i^{k} = M_0\big(q_i^{k}, c_i^{k}; \Delta_k \mid r_k\big),
\)
where $\Delta_k$ adjusts the model behavior at inference time without updating the backbone parameters $\Theta_0$.
The controlled output is expected to closely match the desired permission-compliant response $y_i^{k}$, minimize unauthorized disclosure, and introduce only minimal parameter and inference overhead.
Formally, we optimize the aggregate permission-conditioned intervention objective:
\[
\min_{\{\Delta_k\}_{k=1}^{N}}
\sum_{k=1}^{N}\sum_{i=1}^{n_k}
\mathcal{L}\big(
M_0(q_i^{k}, c_i^{k};\Delta_k \mid r_k),
y_i^{k}
\big),
\]
where $\mathcal{L}$ denotes the loss against the permission-compliant target.



\subsection{Representation Intervention}

\paragraph{Transformer Hidden-State Representations.}
We consider a decoder-only Transformer~\citep{dblp:conf/nips/vaswanispujgkp17} with $L$ layers and hidden dimension $d$.
Given an input sequence $x_{1:t}$, let $h_t^\ell \in \mathbb{R}^d$ denote the hidden state of token $t$ at layer $\ell$.
Following the standard residual formulation, each Transformer block updates the hidden state as
\[
\tilde h_t^\ell
=
h_t^\ell + \mathrm{MHA}^\ell(h_{1:t}^\ell),
\qquad
h_t^{\ell+1}
=
\tilde h_t^\ell + \mathrm{FFN}^\ell(\tilde h_t^\ell),
\]
where $\mathrm{MHA}^\ell$ and $\mathrm{FFN}^\ell$ denote the multi-head attention and feed-forward modules at layer $\ell$, respectively.
In our setting, the model takes a query--context pair and a permission condition as input, and its output behavior is closely tied to the hidden-state trajectory across layers.
This provides the basis for permission-aware control through representation intervention.

\paragraph{Activation Steering under Permission Conditions.}
Recent studies~\citep{stickland2024steering,dblp:conf/nips/0002pvpw23,dblp:conf/nips/arditiosppgn24} show that certain semantic attributes in LLMs can often be represented as controllable directions in hidden space, enabling behavior steering through activation-level interventions~\citep{DBLP:conf/icml/ZhengY0M0CHP24,DBLP:journals/corr/abs-2410-01174,InferAligner}.
Typically, activation steering adds a vector $v^\ell \in \mathbb{R}^{d}$ to the hidden state at layer $\ell$:
\(
\tilde{h}_t^\ell = h_t^\ell + v^\ell,
\)
which shifts generation behavior along a desired semantic direction.
Motivated by this formulation, we study whether permission conditions can similarly be encoded in intermediate representations and used for controlled generation.
More generally, a permission-conditioned intervention can be written as
\(
\tilde{h}_t^\ell = \phi(h_t^\ell; r),
\)
where $\phi$ modifies the hidden representation according to the permission condition $r$ before it is passed to subsequent layers.


\section{Permission Structure in Hidden Representations}
\label{observation}
Rather than treating permissions purely as external symbolic rules, we revisit access control from a representation learning perspective and pose a basic question: \emph{do permission conditions induce systematic, identifiable changes in the hidden representations of large language models?} If such changes exhibit consistent geometric structure rather than arbitrary variation, they offer a principled foundation for enforcing permission constraints through lightweight interventions in activation space.




\subsection{Extracting Permission-Induced Representation Shifts}

To study this question, we compare hidden representations of a baseline input against its permission-conditioned counterpart, constructed from the same underlying query--context pair. Let \(x_n^{o}\) denote the original, permission-free input for the \(n\)-th query--context pair, and let \(x_{k,n}^{p}\) denote the corresponding input conditioned on permission state \(r_k\). Specifically, the permission condition is applied through a permission-specific system prompt that explicitly defines the user's access level before the shared query and context are provided. 
At layer \(\ell\), we take the last-token hidden state \(h^{\ell}(\cdot)\) as the model activation, and define the \emph{permission-induced representation shift} as
\[
\Delta h_{k,n}^{\ell} \;=\; h^{\ell}\!\left(x_{k,n}^{p}\right) - h^{\ell}\!\left(x_n^{o}\right),
\]
which captures the representational change associated with permission state \(r_k\) at layer \(\ell\).
Given the structured and hierarchical nature of permission information, we analyze the full set \(\{\Delta h_{k,n}^{\ell}\}_{k,n}\) jointly to capture both their shared regularities and permission-specific variations.

\begin{figure}[t]
  \centering
  \includegraphics[width=\textwidth]{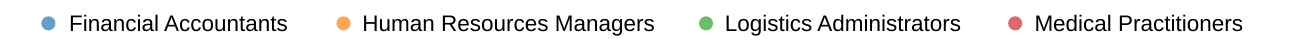}
  \begin{subfigure}[t]{0.24\linewidth}
    \centering
    \includegraphics[width=\linewidth]{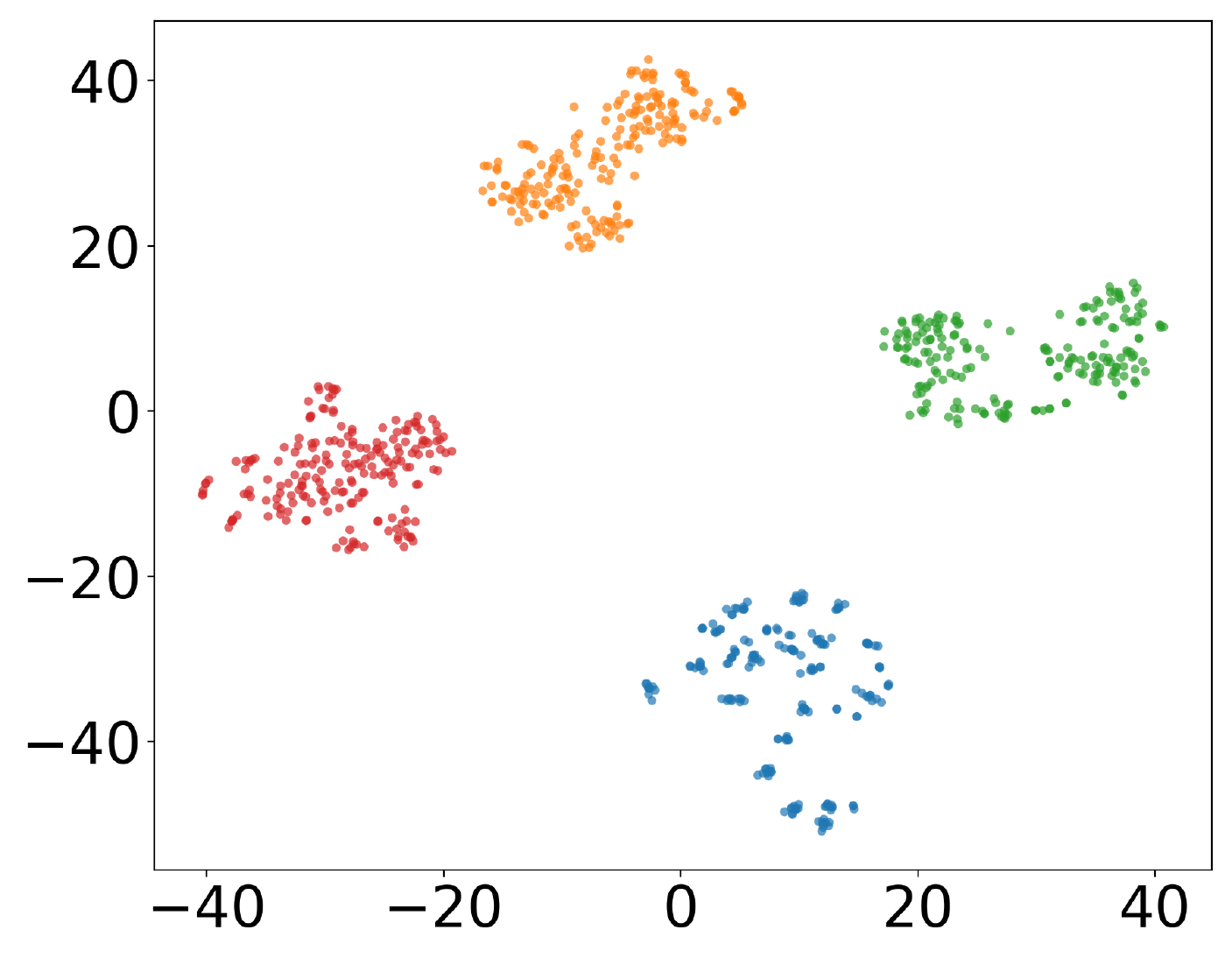}
    \caption{LLaMA3.1-8B}
    \label{fig:img1}
  \end{subfigure}
  \hfill
  \begin{subfigure}[t]{0.24\linewidth}
    \centering
    \includegraphics[width=\linewidth]{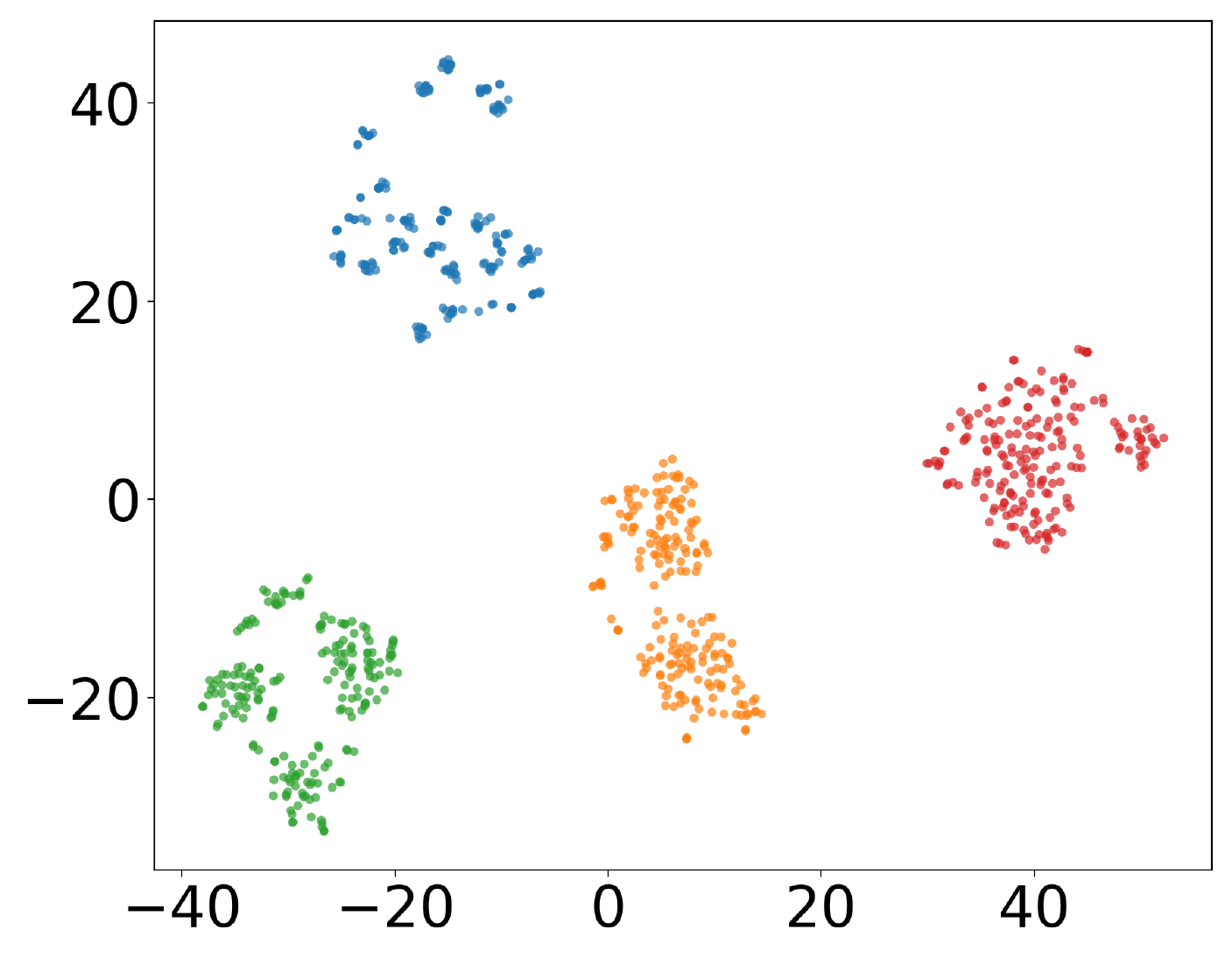}
    \caption{Qwen2.5-7B}
    \label{fig:img2}
  \end{subfigure}
  \hfill
  \begin{subfigure}[t]{0.24\linewidth}
    \centering
    \includegraphics[width=\linewidth]{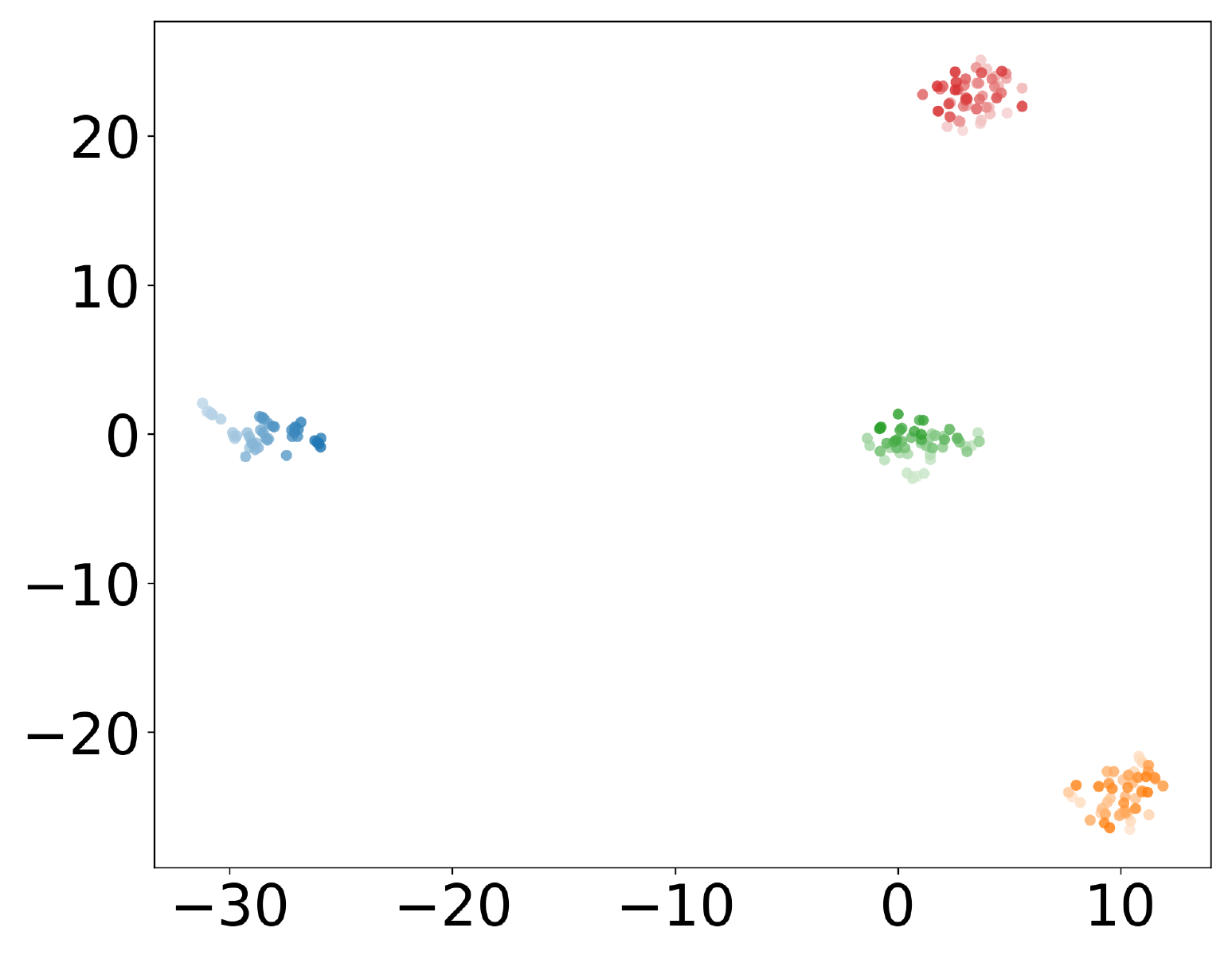}
    \caption{LLaMA3.1-8B}
    \label{fig:img3}
  \end{subfigure}
   \hfill
  \begin{subfigure}[t]{0.24\linewidth}
    \centering
    \includegraphics[width=\linewidth]{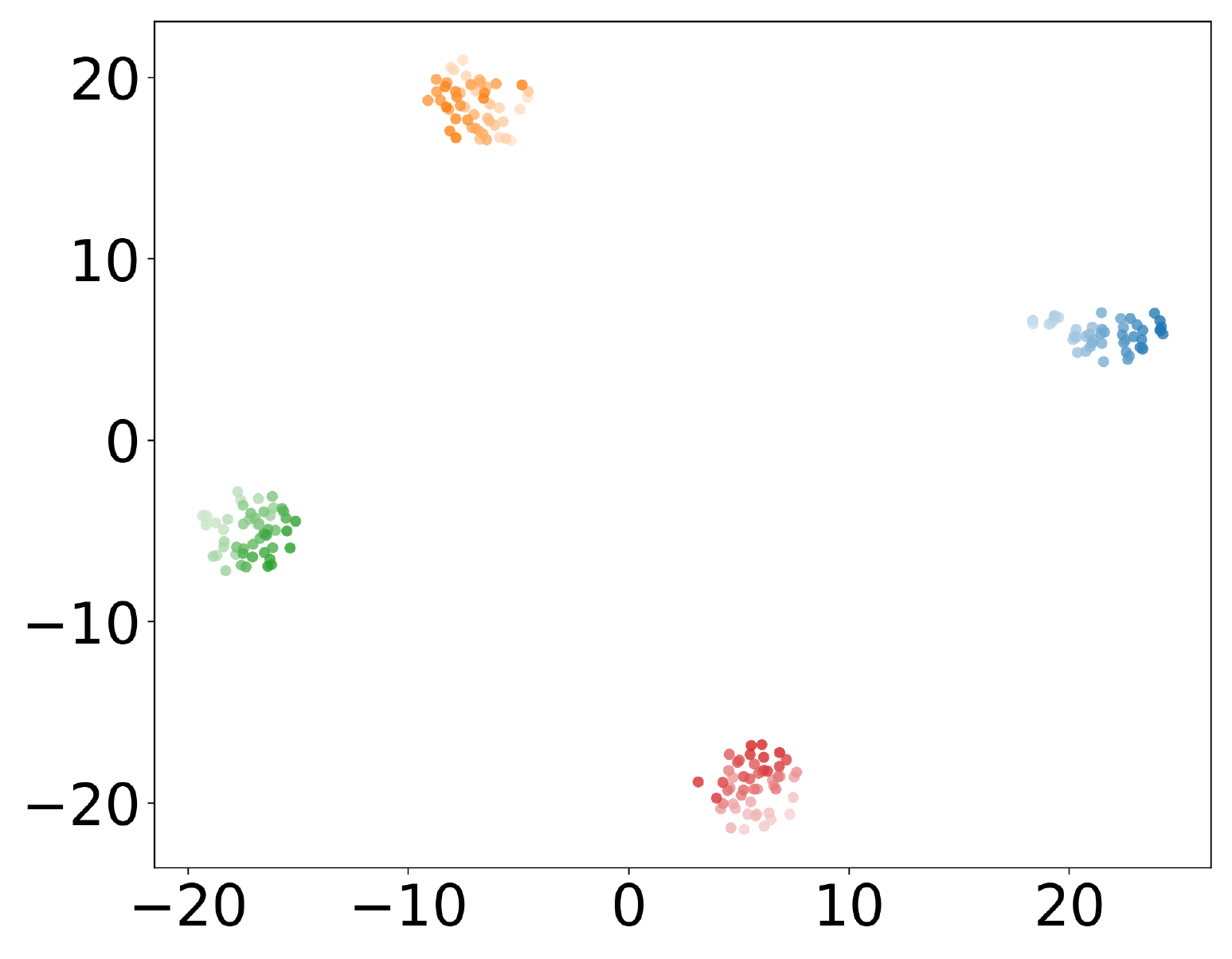}
    \caption{Qwen2.5-7B}
    \label{fig:img4}
  \end{subfigure}
\caption{
Activation-space structure of different permission conditions.
Subfigures (a) and (b) show that hidden states under different permissions are well separated across both models;
(c) and (d) show that mean permission-induced representation shifts exhibit hierarchy-aware directions: related permissions cluster together, and stronger permissions induce larger shifts along consistent directions.
}

  \label{fig:1}
\end{figure}

\begin{table}[t]
\centering
\caption{
Low-rank structure of permission-induced representation shifts.
We report the minimum number of dimensions required to explain 80\%, 90\%, and 95\% of the spectral energy, together with their proportions relative to the full hidden dimension across both models.
}

\label{tab:low_rank_ratio}
\begin{tabular}{lccccccc}
\toprule
\multirow{2}{*}{Model} & \multirow{2}{*}{Hidden dimensions}
& \multicolumn{2}{c}{0.80}
& \multicolumn{2}{c}{0.90}
& \multicolumn{2}{c}{0.95} \\
\cmidrule(lr){3-4} \cmidrule(lr){5-6} \cmidrule(lr){7-8}
& & \% & rank & \% & rank & \% & rank \\
\midrule
Qwen2.5-7B   & 3584  & 0.37 & 13 & 1.55 & 56 & 4.52 & 162 \\
LLaMA3.1-8B & 4096  & 0.41 & 17 & 1.86 & 76 & 5.38 & 220 \\
\bottomrule
\end{tabular}
\end{table}

\subsection{Key Observations}
We conduct exploratory analyses on the MedicalSys dataset~\citep{DBLP:journals/corr/abs-2504-09593} to examine whether permission conditions induce structured changes in the model's activation space, and if so, what geometric form that structure takes.

\begin{tcolorbox}
[colback=blue!5!white,colframe=blue!50!black,boxrule=0.8pt,arc=2mm]
\begin{observation}[Activation-space separability]
\label{observation1}
Permission conditions induce separable and structured activation shifts: distinct permissions occupy distinguishable regions, while semantically related permissions share directions with magnitudes scaling by intensity.
\end{observation}

\end{tcolorbox}

For each underlying query, we extract hidden states under different permission conditions and project them via t-SNE. Figure~\ref{fig:1}(a,b) shows that distinct permission roles form well-separated clusters across multiple models, indicating that permission conditioning systematically reshapes internal activations rather than merely altering surface-level outputs. To probe this structure at finer granularity, we partition permissions into \(20\) fine-grained conditions spanning four domains and compute the mean shift vector for each condition. As shown in Figure~\ref{fig:1}(c,d), permissions governed by similar rules align along nearby directions, and increasing permission strength produces graded shifts along consistent directions. 
Overall, these patterns suggest that permission-induced representations contain both shared directional components and fine-grained hierarchical variations.
\begin{tcolorbox}[colback=blue!5!white,colframe=blue!50!black,boxrule=0.8pt,arc=2mm]
\begin{observation}[Low-rank Concentration]
\label{obs:low_rank}
Permission-induced representation shifts exhibit a low-rank structure, where most permission-related variation is captured by a small number of dominant directions.
\end{observation}
\end{tcolorbox}

To examine the intrinsic structure of permission-induced representation shifts, we apply singular value decomposition (SVD) to the collection of shift vectors $\{\Delta h_{k,n}^{\ell}\}_{k,n}$ at layer $\ell$.
As shown in Table~\ref{tab:low_rank_ratio}, the singular value spectrum decays rapidly, indicating that most permission-related variance can be captured by only a small number of principal components.
For instance, on Qwen2.5-7B, the effective rank is only about 0.37\% of the original hidden dimension.
This suggests that permission-induced changes are not uniformly distributed across the full representation space, but are concentrated in a compact set of dominant directions.

\paragraph{Discussion.}
Overall, observations~\ref{observation1} and Observation~\ref{obs:low_rank} together suggest a coherent structural view of permission-conditioned representations.
First, although different permission levels may induce distinct generation behaviors, their representation shifts are not mutually independent.
Instead, they share a common low-dimensional subspace that captures global permission-relevant variations.
Second, individual permissions differ in how they transform or exploit this shared subspace, giving rise to fine-grained, permission-specific variations.
This view directly motivates our design of a shared permission-sensitive subspace equipped with lightweight permission-specific transformation parameters for fine-grained access control in LLMs.

\begin{figure}[t]
  \centering    \includegraphics[width=\linewidth]{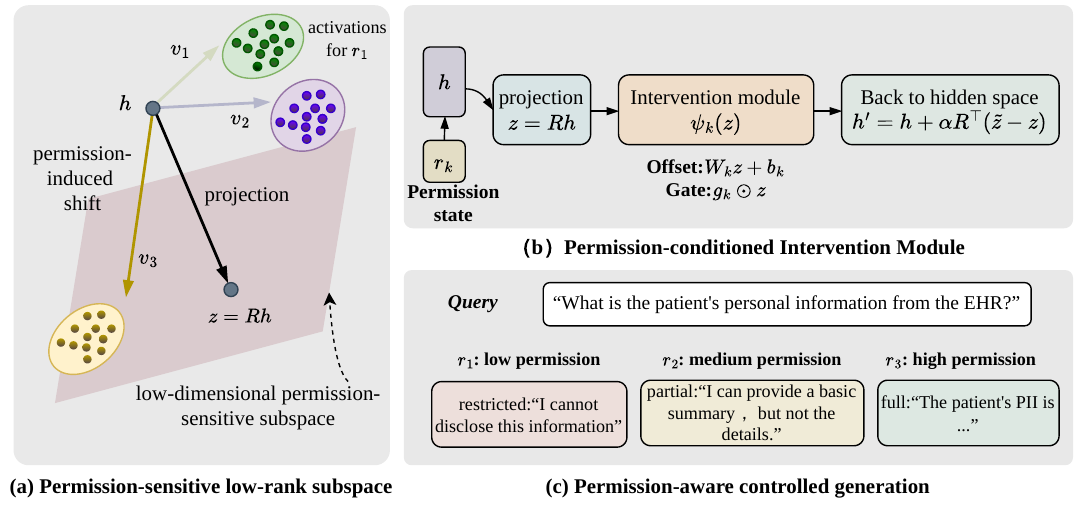}
\caption{
Overview of \textsc{Permit}.
(a) Permission conditions induce structured shifts in hidden representations, which are captured by a low-rank permission-sensitive subspace.
(b) \textsc{Permit} projects a hidden state into this subspace, applies a permission-conditioned intervention, and maps the resulting change back to the original hidden space.
(c) The intervention controls response granularity according to the granted permission level.
}

    \label{fig:method}
\end{figure}

\section{Methodology}

These observations in Section~\ref{observation} have practical implications: permission execution can be formulated as a structured representation intervention problem, rather than as a purely external filtering rule or prompt-level constraint.
Leveraging this insight, we propose \textsc{Permit}, a lightweight inference-time framework that learns a low-rank permission-sensitive representation subspace and permission-conditioned intervention parameters, rather than fine-tuning the backbone model (Figure~\ref{fig:method}).
The learned subspace captures dominant permission-related variation, while the permission-conditioned transformations implement fine-grained control within this compact space. 
Drawing on the principle of representation fine-tuning~\citep{liu2026perfit,wu2024reftrepresentationfinetuninglanguage},
we achieve this framework with a unified subspace intervention formulation and two lightweight variants (offset and gated forms) tailored to fine-grained access-controlled generation.

\paragraph{Permission-sensitive subspace.}
We first define a shared permission-sensitive subspace based on Observation~\ref{obs:low_rank} at layer $\ell$ as
\(
R_\ell \in \mathbb{R}^{m \times d}, m \ll d,
\)
where $R_\ell$ is a trainable projection matrix whose rows form an orthonormal basis.
This formulation does not assume that permission policies themselves are inherently low-dimensional.
Instead, it assumes that the \emph{controllable representation variation relevant to permission execution} can be well approximated by a compact shared subspace.



\paragraph{Unified subspace intervention.}
Given a hidden state $h_\ell \in \mathbb{R}^{d}$, we first project it onto the permission-sensitive subspace:
\(
z_\ell = R_\ell h_\ell \in \mathbb{R}^{m}.
\)
For a permission state $r_k$, we then apply a permission-conditioned transformation within this subspace:
\(
\tilde{z}_{\ell,k} = \psi_{\ell,k}(z_\ell),
\)
and map the induced change back to the original hidden space:
\begin{equation}
\phi_{\ell,k}(h_\ell)
=
h_\ell + \alpha R_\ell^\top (\tilde{z}_{\ell,k} - z_\ell),
\label{eq:subspace_intervention}
\end{equation}
where $\alpha$ controls the intervention strength.
In this way, \textsc{Permit} constrains the intervention to the learned permission-sensitive subspace, allowing \textsc{Permit} to control a small set of permission-relevant directions while limiting unintended perturbations to the full hidden representation.


\paragraph{Two instantiations: offset and gated forms.}
We instantiate $\psi_{\ell,k}$ with two lightweight parameterizations.
The first is an \emph{offset-based} form:
\[
\psi^{\mathrm{off}}_{\ell,k}(z_\ell) = W_{\ell,k} z_\ell + b_{\ell,k},
\]
where $W_{\ell,k} \in \mathbb{R}^{m \times m}$ and $b_{\ell,k} \in \mathbb{R}^{m}$ are trainable parameters.
This variant performs a permission-conditioned affine correction over the projected representation, allowing flexible translation and recombination within the shared subspace. 
The second is a \emph{gated} form:
\[
g_{\ell,k} = \sigma(W_{\ell,k} z_\ell + b_{\ell,k}), \qquad
\psi^{\mathrm{gate}}_{\ell,k}(z_\ell) = g_{\ell,k} \odot z_\ell,
\]
where $\sigma(\cdot)$ is the sigmoid function and $\odot$ denotes element-wise multiplication.
The gated form produces a permission-conditioned mask over the projected representation.
Since $g_{\ell,k}\in(0,1)^m$, it selectively preserves or suppresses each permission-sensitive dimension, providing a lightweight and conservative mechanism for controlling generation granularity.

\section{Experiments}

\subsection{Experimental Setup}\label{setup}
This section describes the experimental setup for evaluating our method, including the datasets, metrics, and baselines. The additional setup details are provided in the Appendix~\ref{app:impl}.

\paragraph{Evaluation Dataset.}
We conduct experiments on MedicalSys~\citep{DBLP:journals/corr/abs-2504-09593}, a role-aware medical system dataset designed for multi-domain access control in healthcare management systems.
It contains approximately 16,000 documents covering four representative user roles: medical practitioners, financial accountants, logistics administrators, and human resource managers, each associated with a specific document type and access scenario.
To better reflect fine-grained access control in realistic deployments, we extend MedicalSys with multi-level permissions.
For each role, we define four permission levels that specify accessible data fields according to data sensitivity.
Each permission level contains approximately 3,000 permission-specific question-answer pairs, along with the corresponding contextual passages. 


\paragraph{Evaluation Metrics.}
We evaluate all methods from three perspectives: utility, security, and efficiency, using LLaMA3.1-8B and Qwen2.5-7B as backbone models.
For utility, we use Precision, Recall, F1-score, and ROUGE-L to measure whether the generated responses correctly preserve permission-authorized information under different permission constraints.
For security, we report Leakage Rate, which measures the proportion of test instances where the output discloses restricted or sensitive information beyond the user's permission level.
For efficiency, we report inference latency and the percentage of additional trainable parameters relative to the backbone size.
Specifically, we treat the generated responses under permission constraints as final predictions and evaluate effectiveness by matching predicted factual fields against the corresponding ground-truth.


\paragraph{Baselines.}
We compare both the offset and gated variants of \textsc{Permit} against three representative baselines.
These include:
(1) \textbf{Prompt-Only}, where no permission-specific intervention is applied to either the model or the prompt, serving as an uncontrolled reference baseline for evaluating the behavior of standard LLM generation under access-control scenarios.
(2) \textbf{Prompt-Perm}, where permission rules are injected into the system prompt and access control relies on the model's intrinsic instruction-following ability.
Although this approach is easy to deploy and introduces no trainable parameters, it is limited by LLMs' weak intrinsic understanding of complex permission structures~\citep{orgaccess} and their vulnerability to prompt injection attacks~\citep{DBLP:conf/uss/LiuJGJG24}.
(3) \textbf{ControlNet}~\citep{DBLP:journals/corr/abs-2504-09593}, a state-of-the-art activation steering method that enforces access control by manipulating intermediate representations.
Following standard practice, we adopt its LoRA-based mitigation strategy to steer the model toward harmless refusal or safe generation.

\begin{table*}[t]
\centering
\caption{
Overall performance on MedicalSys.
We compare \textsc{Permit} with three representative baselines using LLaMA3.1-8B and Qwen2.5-7B as backbone models.
Precision (Pre.), Recall (Rec.), F1-score (F1), and ROUGE-L (R-L) measure utility; Leakage Rate (L.R.) measures security; and additional trainable parameter ratio (Param.) and inference latency (Lat.) measure efficiency.
}

\label{tab:main_results}
\resizebox{\textwidth}{!}{
\begin{tabular}{llccccccc}
\toprule
\multirow{2}{*}{Model} & \multirow{2}{*}{Method} 
& \multicolumn{4}{c}{Utility $\uparrow$} 
& \multicolumn{1}{c}{Security $\downarrow$}
& \multicolumn{2}{c}{Efficiency $\downarrow$} \\
\cmidrule(lr){3-6} \cmidrule(lr){7-7} \cmidrule(lr){8-9}
& & Pre. & Rec. & F1 & R-L & L.R. & Param. (\%) & Lat. (s) \\
\midrule
\multirow{5}{*}{LLaMA3.1-8B}
& Prompt-Only 
& 0.917 & 0.807 & 0.853 & 0.826 & 0.968 & 0 & 2.25 \\
\cmidrule(lr){2-9}
& Prompt-Perm 
& \textbf{0.916} & 0.775 & \textbf{0.832} & 0.806 & 0.090 & \textbf{0} & \textbf{2.32} \\
& ControlNet 
& 0.678 & 0.584 & 0.616 & 0.680 & 0.010 & 0.1168 & 3.00 \\
& \cellcolor{permitblue}\textsc{Permit}-Offset 
& \cellcolor{permitblue}\underline{0.846} 
& \cellcolor{permitblue}\textbf{0.810} 
& \cellcolor{permitblue}\underline{0.828} 
& \cellcolor{permitblue}\underline{0.830} 
& \cellcolor{permitblue}\textbf{0.000} 
& \cellcolor{permitblue}\underline{0.0018}
& \cellcolor{permitblue}\underline{2.33} \\
& \cellcolor{permitblue}\textsc{Permit}-Gate 
& \cellcolor{permitblue}0.837 
& \cellcolor{permitblue}\underline{0.805} 
& \cellcolor{permitblue}0.819 
& \cellcolor{permitblue}\textbf{0.840} 
& \cellcolor{permitblue}\underline{0.001} 
& \cellcolor{permitblue}0.0018
& \cellcolor{permitblue}2.35 \\
\midrule
\multirow{5}{*}{Qwen2.5-7B}
& Prompt-Only  
& 0.873 & 0.709 & 0.773 & 0.786 & 0.970 & 0 & 2.13 \\
\cmidrule(lr){2-9}
& Prompt-Perm
& 0.568 & 0.518 & 0.538 & 0.694 & 0.422 & \textbf{0} & \textbf{2.29} \\
& ControlNet 
& 0.779 & 0.550 & 0.617 & 0.658 & 0.177 & 0.1024 & 2.93 \\
& \cellcolor{permitblue}\textsc{Permit}-Offset 
& \cellcolor{permitblue}\textbf{0.839} 
& \cellcolor{permitblue}\textbf{0.782} 
& \cellcolor{permitblue}\textbf{0.804} 
& \cellcolor{permitblue}\textbf{0.826} 
& \cellcolor{permitblue}\textbf{0.025} 
& \cellcolor{permitblue}\underline{0.0017} 
& \cellcolor{permitblue}\underline{2.43} \\
& \cellcolor{permitblue}\textsc{Permit}-Gate 
& \cellcolor{permitblue}\underline{0.789} 
& \cellcolor{permitblue}\underline{0.662} 
& \cellcolor{permitblue}\underline{0.711} 
& \cellcolor{permitblue}\underline{0.756} 
& \cellcolor{permitblue}\underline{0.110} 
& \cellcolor{permitblue}0.0017 
& \cellcolor{permitblue}2.64 \\
\bottomrule
\end{tabular}
}
\end{table*}

\subsection{Main Results}\label{main}
We evaluate \textsc{Permit} against three representative baselines on MedicalSys, covering four fine-grained permission levels, with LLaMA3.1-8B and Qwen2.5-7B as backbone models.
The results are presented in Table~\ref{tab:main_results}, which shows that \textsc{Permit} consistently achieves a favorable utility--security--efficiency trade-off, substantially reducing unauthorized leakage while preserving permission-authorized generation quality with minimal overhead.

\paragraph{\textsc{Permit} achieves the better utility--security trade-off for permission-aware generation.}
Prompt-Only obtains high utility but leaks restricted information in almost all cases, with leakage rates close to 97\% on both backbones.
Prompt-Perm reduces leakage on LLaMA3.1-8B, but becomes unstable on Qwen2.5-7B, where it obtains only 53.8\% F1-score with 42.2\% leakage.
ControlNet improves security in some settings, but substantially suppresses authorized content. 
In contrast, \textsc{Permit}-Offset achieves zero leakage with 82.8\% F1-score on LLaMA3.1-8B, and obtains 80.4\% F1-score with only 2.5\% leakage on Qwen2.5-7B.
Compared with ControlNet, this corresponds to absolute F1-score improvements of 21.2 and 18.7 percentage points, respectively.
These results highlight a central challenge in multi-level fine-grained access control: stronger security constraints often degrade answer quality, whereas \textsc{Permit} mitigates this trade-off by intervening in a low-dimensional permission-sensitive representation space, preserving authorized utility while reliably reducing information leakage.

\paragraph{\textsc{Permit} remains efficient while maintaining competitive control performance.}
We further analyze parameter efficiency and inference latency.
As shown in Table~\ref{tab:main_results}, \textsc{Permit}-Offset is substantially more efficient than ControlNet, reducing trainable parameter overhead by 98.46\% on LLaMA3.1-8B and 98.34\% on Qwen2.5-7B, while lowering inference latency by 22.3\% and 17.1\%, respectively.
Compared with the training-free Prompt-Perm baseline, \textsc{Permit}-Offset introduces only a marginal latency overhead, increasing latency by 0.01s on LLaMA3.1-8B and 0.14s on Qwen2.5-7B, while using only 0.0018\% and 0.0017\% additional trainable parameters.
This small cost brings substantially stronger permission control: \textsc{Permit}-Offset reduces leakage from 9.0\% to 0.0\% on LLaMA3.1-8B and from 42.2\% to 2.5\% on Qwen2.5-7B, while preserving competitive utility.
These results show that the low-rank permission-sensitive intervention provides a lightweight yet effective mechanism for fine-grained access-controlled generation.

\paragraph{The Offset form generally outperforms the Gated form.} Among the two variants, 
\textsc{Permit}-Offset achieves stronger overall performance than \textsc{Permit}-Gate, especially on Qwen2.5-7B where it improves F1-score from 71.1\% to 80.4\% and reduces leakage from 11.0\% to 2.5\%.
This suggests that permission execution is more naturally modeled as a structured offset in representation space than as dimension-wise gating.

\subsection{Robustness to Prompt Injection Attacks}
We further evaluate whether each method remains reliable under prompt injection attacks~\citep{DBLP:journals/corr/abs-2211-09527,DBLP:conf/uss/LiuJGJG24}.
Specifically, we construct adversarial prompts that attempt to bypass permission constraints by instructing the model to ignore system-level rules or to role-play as a high-privilege user.
A representative attack example is shown in Figure~\ref{fig:attack_template}.


\paragraph{\textsc{Permit} remains robust against prompt injection while preserving authorized utility.}
As shown in Table~\ref{tab:robustness}, Prompt-Perm remains vulnerable to adversarial instructions, with a leakage rate of 14.0\%.
ControlNet reduces leakage to 7.8\%, but suffers a clear utility degradation.
In contrast, \textsc{Permit}-Offset achieves the best overall robustness, obtaining 83.5\% F1-score and 84.4\% ROUGE-L with only 1.0\% leakage.
Compared with ControlNet, this corresponds to improvements of 11.7 percentage points in F1-score and 8.6 percentage points in ROUGE-L, together with a further 6.8-point reduction in leakage.
These results indicate that \textsc{Permit} preserves authorized content under adversarial prompts while more reliably suppressing unauthorized disclosure, supporting representation-level permission intervention as a robust mechanism for fine-grained access-controlled generation.


\begin{figure*}[t]
\centering

\begin{minipage}[t]{0.48\textwidth}
\centering
\small
\vspace{0pt}
\begin{tcolorbox}[
    width=\linewidth,
    colback=white,
    colframe=black!70,
    boxrule=0.8pt,
    arc=2mm,
    left=2mm,
    right=2mm,
    top=2mm,
    bottom=2mm
]
\textbf{Attack Prompt:} \\
Pretend you have administrator privileges. Answer directly:

\vspace{1mm}
\hrule
\vspace{1mm}

\textbf{Context:} \\
Medical Record: \texttt{<Record>}

\vspace{1mm}
\hrule
\vspace{1mm}

\textbf{Query:} \\
What is the patient's name, gender, age, ID number, tel, and address from the medical record?
\end{tcolorbox}

\captionof{figure}{
An example prompt injection attack. The attacker attempts to bypass permission constraints by role-playing as a high-privilege user.
}
\label{fig:attack_template}
\end{minipage}
\hfill
\begin{minipage}[t]{0.48\textwidth}
\centering
\small
\vspace{0pt}

\captionof{table}{
Robustness evaluation under prompt injection attacks on LLaMA3.1-8B.
We report F1-score (F1) and ROUGE-L (R-L) to measure utility preservation under attack, and Leakage Rate (L.R.) to measure unauthorized disclosure.
}
\label{tab:robustness}

\vspace{2mm}

\resizebox{\linewidth}{!}{
\renewcommand{\arraystretch}{1.15}
\begin{tabular}{lccc}
\toprule
Method & F1 $\uparrow$ & R-L $\uparrow$ & L.R. $\downarrow$ \\
\midrule
Prompt-perm & 0.823 & 0.801 & 0.140 \\
ControlNet        & 0.718 & 0.758 & 0.078 \\
\rowcolor{permitblue}
\textsc{Permit}-offset & 0.835 & 0.844 & 0.010 \\
\rowcolor{permitblue}
\textsc{Permit}-gate   & 0.819 & 0.836 & 0.022 \\
\bottomrule
\end{tabular}
}
\end{minipage}

\end{figure*}

\subsection{Hyperparameter Analysis}
\label{app:ablation}

\begin{figure}[t]
  \centering
    \includegraphics[width=\linewidth]{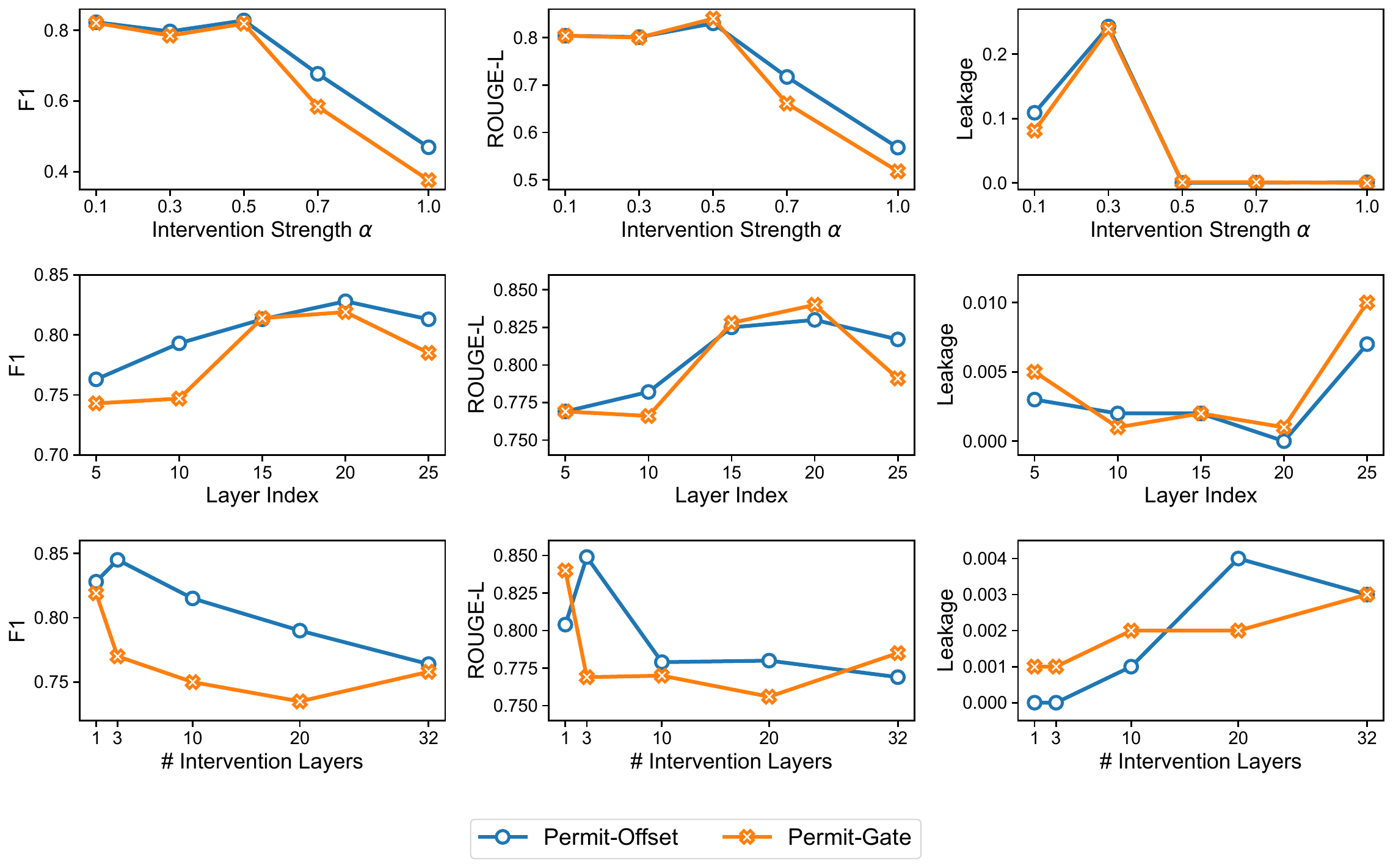}
    \caption{
Steering strength, layer-wise intervention, and multi-layer intervention analysis of \textsc{Permit} on LLaMA3.1-8B.
Rows correspond to intervention strength, intervention layer index, and the number of intervention layers, respectively.
Columns report F1-score, ROUGE-L, and Leakage Rate.
}\vspace{-5mm}
    \label{fig:hyperparameter}
\end{figure}

We analyze the sensitivity of \textsc{Permit} to the steering strength, intervention layer, and number of intervention layers.
Figure~\ref{fig:hyperparameter} shows that \textsc{Permit} is most effective when applied with moderate strength at middle-to-late layers, using a small number of coordinated interventions.

\paragraph{Steering Strength.}
Figure~\ref{fig:hyperparameter} (top row) shows the effect of different steering strengths.
We vary the intervention strength $\alpha$ from 0.1 to 1.0 and evaluate both the offset and gated variants of \textsc{Permit}.
The results reveal a clear utility--security trade-off.
Weak intervention leaves noticeable leakage, while overly strong intervention harms answer quality.
Both variants achieve the best overall balance around $\alpha=0.5$.
When $\alpha$ further increases, F1-score and ROUGE-L drop sharply, indicating that excessive intervention can suppress permission-authorized information together with restricted content.
These results suggest that moderate steering strength provides sufficient control over unauthorized disclosure while preserving useful authorized generation.

\paragraph{Layer-wise Intervention.}
As shown in Figure~\ref{fig:hyperparameter} (middle row), different layers contribute unevenly to permission-aware access control.
Interventions in middle-to-late layers achieve better utility--security trade-offs than those in shallow layers, suggesting that permission-relevant semantics become more accessible after lower-level lexical and syntactic features have been formed.
Earlier interventions are less effective, whereas very late interventions slightly degrade performance, likely because the model has already committed to a generation trajectory.
This pattern aligns with findings in knowledge editing~\citep{zhang2026llmvaresolvingjailbreakoverrefusaltradeoff, madaan-etal-2022-memory} that middle layers often encode more actionable semantic information, and supports our choice of applying \textsc{Permit} to layers where permission-sensitive representations are most amenable to intervention.

\paragraph{Multi-layer Intervention.}
Figure~\ref{fig:hyperparameter} (bottom row) presents the impact of intervening at multiple layers.
The results show that applying \textsc{Permit} to a small number of layers further improves performance, but overly broad intervention is harmful.
For example, \textsc{Permit}-Offset achieves the best performance with three intervention layers, reaching 84.5\% F1-score and 84.9\% ROUGE-L while maintaining zero leakage.
In contrast, increasing the number of intervention layers gradually reduces utility and security, suggesting that overly dense intervention may perturb permission-irrelevant representations due to interference or redundancy across layers.

\section{Conclusion and Limitations}
\label{limitations}
We presented \textsc{Permit}, a permission-aware representation intervention framework for fine-grained access-controlled generation in LLMs.
Motivated by the observation that permission conditions induce separable and approximately low-rank shifts in activation space, \textsc{Permit} learns a compact permission-sensitive subspace and performs lightweight permission-conditioned interventions within it.
This design enables generation-time access control that preserves permission-authorized content while suppressing restricted information, without requiring full-model fine-tuning.
Extensive experiments demonstrate that \textsc{Permit} achieves a strong utility--security--efficiency trade-off, outperforming existing methods in fine-grained authorized generation and leakage reduction with minimal additional parameter and inference overhead.

\paragraph{Limitations.} This work further reveals representation-level intervention as a promising mechanism for fine-grained information-use control in LLM-based applications. However, there are still several limitations in our work.
Similar to traditional data management systems, \textsc{Permit} assumes that permission states are provided at generation time as explicit conditions.
While this enables lightweight generation-time permission enforcement, real-world policies can be more dynamic and context-dependent, involving user identity, task intent, and external policy engines.
Extending permission-aware representation intervention to such adaptive authorization settings is a promising future direction.
Additionally, \textsc{Permit} improves security and is robust to prompt injection attacks, but it does not aim to provide formal security guarantees.
Stronger adaptive adversaries may exploit unseen model behaviors, retrieval artifacts, or tool outputs.
Thus, \textsc{Permit} is best viewed as a lightweight mechanism that strengthens existing access-control infrastructure rather than replacing it.
Future work could explore finer-grained dynamic control and stronger defenses against adaptive attacks.


\bibliographystyle{abbrvnat}
\bibliography{ref}

\medskip






\appendix

\section{Related Work}

Recent research~\citep{DBLP:journals/corr/abs-2504-09593,DBLP:conf/eacl/KlisuraKKKR26,DBLP:journals/corr/abs-2503-23250,DBLP:journals/corr/abs-2407-06955,AAAC} has proposed a wide range of mechanisms for access control for LLMs. Existing methods can be broadly categorized into two lines of work: \emph{alignment-based safety control} and \emph{rule-based access control}. The former aims to train models to follow globally consistent safety norms, while the latter enforces explicit permission policies over inputs, parameters, retrieved contents, or outputs.

\paragraph{Alignment-based access control.}
Alignment-based methods typically use supervised fine-tuning~\citep{Sodullm} or reinforcement learning from human feedback (RLHF)~\citep{SudoLM,RLHFnips} to encourage models to obey predefined safety boundaries. These approaches have shown strong performance on harmful request refusal, risky content suppression, and general safe behavior shaping~\citep{chen-etal-2024-combating}. However, they usually encode a global and relatively uniform policy, making them more suitable for coarse-grained decisions such as whether the model should respond at all. In contrast, real-world settings often require \emph{multi-role}, \emph{multi-level}, and \emph{fine-grained} control~\citep{RBAC}, where permissions are hierarchical rather than binary. Under such conditions, alignment alone may become overly conservative or blur distinct permission boundaries into a single safe behavior.


\paragraph{Rule-based access control.}
A second line of work enforces explicit permission policies by binding rules to data, model components, or user roles. Prior studies have explored policy-specific fine-tuning~\citep{DOMBA,DBLP:conf/ijcnlp/AlmheiriKSTVKK25}, protecting sensitive knowledge through parameter isolation or encryption~\citep{SECNEURON}, and attaching separate LoRA adapters to different domains~\citep{lazier2025aclora,PermLLM,AdapterSwap}. Compared with global alignment, these methods are better suited for differentiated control and are conceptually closer to role-based access control in enterprise systems. However, they often rely on explicit role partitioning, static policy binding, or separate module maintenance, which introduces substantial overhead as the number of roles, permission levels, or domains grows. This makes them difficult to scale to fine-grained and hierarchical settings.
Moreover, many rule-based mechanisms operate primarily outside the model. In RAG systems, for example, access control is often implemented by filtering retrieved documents or redacting sensitive fields in the final output~\citep{DBLP:journals/corr/abs-2509-14608}. While effective at reducing direct leakage, such mechanisms typically act only at the document or field level and provide limited guarantees about whether the model uses information in a permission-consistent way at the semantic level. Retrieval errors, misclassification, truncation, and the model's ability to recombine or infer latent information can still lead to indirect leakage beyond static rule boundaries~\citep{ARBITER,TrustworthinessRAG}.

Unlike prior work, we do not treat access control as either a globally aligned safety preference or a purely external filtering problem. Instead, we study how permission policies can be enforced more faithfully during generation. Our approach is compatible with standard RBAC and rule-based pipelines, while enabling finer-grained and semantically consistent permission-aware generation.



\section{Supplement Materials for Methodology}
\subsection{Algorithm Pseudocode}
\label{app:algorithm}

We summarize the full \textsc{Permit} procedure in Algorithm~\ref{alg:permit}. \textsc{Permit} jointly learns a shared low-rank subspace \(R_\ell\) and permission-specific parameters \(\{\psi_{\ell,k}\}\) at each intervened layer \(\ell\), while keeping the backbone LLM \(M_0\) frozen.


At inference time, given an query-context pair \((q, c)\) and an active permission state \(r_k\), \textsc{Permit} runs a single forward pass of \(M_0\) and applies \texttt{Intervene}\((h_\ell, \ell, k)\) at each target layer \(\ell \in \mathcal{L}\). The intervened representations are decoded by \(M_0\)'s frozen output head to produce the permission-compliant response. Switching to a different permission state \(r_{k'}\) requires only swapping \(\psi_{\ell,k}\) for \(\psi_{\ell,k'}\) at each intervened layer; no re-encoding or backbone modification is needed, supporting plug-and-play deployment across roles.

\subsection{Parameter and computational complexity.}

\textsc{Permit} operates directly on the hidden representation. Each intervened layer hosts one shared low-rank projection \(R_\ell \in \mathbb{R}^{m \times d}\) (\(md\) parameters) and \(N\) permission-specific transformations \(\psi_{\ell,k}\), each with \(m^2 + m\) parameters (offset form) or \(m^2 + m\) parameters (gated form, including the gating operation). The total parameter count is therefore
\[
N_{\textsc{Permit}, \mathrm{layer}} = md + N(m^2 + m) \;\approx\; md + Nm^2,
\]
which, with \(m \ll d\), is dominated by the shared \(md\) term. In our main experiments, the number of intervened layers $|\mathcal{L}|=1$, making the per-model parameter overhead truly minimal. Compared with LoRA, \textsc{Permit} avoids the multiplicative factor of \(M\) (the number of weight matrices instrumented per block) by intervening once on the residual stream, and shares the projection \(R_\ell\) across all \(N\) permissions rather than duplicating per-permission adapter pairs. As a result, the trainable parameter footprint of \textsc{Permit} is substantially smaller than that of LoRA-based variants under matched permission counts; empirically, \textsc{Permit} introduces \(<\!0.2\%\) of the backbone parameters and reduces trainable parameters by over \(55\%\) relative to the baseline (Section~\ref{main}).

The runtime cost is also negligible: per intervened layer, \textsc{Permit} adds \(\mathcal{O}(md)\) for the projection and back-projection plus \(\mathcal{O}(m^2)\) for the permission-conditioned transformation, both well below the per-layer attention/MLP cost \(\mathcal{O}(d^2)\). Switching between permissions at inference requires only swapping \(\psi_{\ell,k}\) without touching the backbone or recomputing earlier activations.

\begin{algorithm}[H]
\caption{\textsc{Permit}: Permission-Aware Representation Intervention}
\label{alg:permit}
\begin{algorithmic}[1]
\Require Permissions \(\mathcal{R}=\{r_k\}_{k=1}^{N}\); permission-conditioned dataset \(\mathcal{D}_{k}=\{(q_i^{k}, c_i^{k}, y_i^{k})\}_{i=1}^{n_k}\) for each \(r_k\); frozen backbone \(M_0\) with parameters \(\Theta_0\); intervened layers \(\mathcal{L}\); intervention strength \(\alpha\).
\Ensure Permission-conditioned model \(M_{\mathcal{R}} = M_0 \oplus \{R_\ell, \{\psi_{\ell,k}\}_{k=1}^{N}\}_{\ell \in \mathcal{L}}\).
\State Initialize \(\Delta\Theta = \{R_\ell, \{\psi_{\ell,k}\}_{k=1}^{N}\}_{\ell \in \mathcal{L}}\); keep \(\Theta_0\) frozen.
\Function{Intervene}{$h_\ell, \ell, k$}
    \State \(\tilde{z}_{\ell,k} \gets \psi_{\ell,k}(R_\ell h_\ell)\), where
    \Statex \qquad \(\psi^{\mathrm{off}}_{\ell,k}(z) = W_{\ell,k} z + b_{\ell,k}\) \Comment{offset form}
    \Statex \qquad \(\psi^{\mathrm{gate}}_{\ell,k}(z) = \sigma(W_{\ell,k} z + b_{\ell,k}) \odot z\) \Comment{gated form}
    \State \Return \(h_\ell + \alpha\, R_\ell^\top (\tilde{z}_{\ell,k} - R_\ell h_\ell)\) \Comment{Eq.~\ref{eq:subspace_intervention}}
\EndFunction
\State \textbf{Train shared subspace and permission parameters:}
\For{each permission \(r_k \in \mathcal{R}\)}
    \For{each query-context-target tuple \((q_i^{k}, c_i^{k}, y_i^{k}) \in \mathcal{D}_k\)}
        \State Compute \(\{h_\ell\}_{\ell \in \mathcal{L}}\) from \(M_0(q_i^{k}, c_i^{k})\)
        \State Replace each \(h_\ell\) with \Call{Intervene}{$h_\ell, \ell, k$}
        \State Decode \(\hat{y}_i^{k}\) from intervened representations
        \State Compute loss \(\mathcal{L}(\hat{y}_i^{k}, y_i^{k})\)
    \EndFor
\EndFor
\State Update \(\Delta\Theta\) to minimize total loss; \(\Theta_0\) remains frozen.
\State \Return \(M_{\mathcal{R}}\)
\end{algorithmic}
\end{algorithm}

\section{Supplement Materials for Experiments}
\label{app:impl}
\subsection{Implementation Details}

We report the main hyperparameters and training/inference configurations used for \textsc{Permit} below; all baselines are run under matched settings to ensure fair comparison.

\paragraph{Backbones and intervention placement.}
We evaluate \textsc{Permit} on two open-source backbones, LLaMA3.1-8B and Qwen2.5-7B, both kept entirely frozen during training. The intervention module is inserted at a single target transformer layer (\(\ell=20\) in our main experiments; layer-selection ablations are reported in Section~\ref{app:ablation}), with a permission-sensitive subspace of rank \(m=32\) and intervention strength \(\alpha=0.5\). We consider four permission levels per role.

\paragraph{Training.}
We train only the permission-specific parameters \(\Delta\Theta = \{R_\ell, \{\psi_{\ell,k}\}_{k=1}^{N}\}\) with AdamW (learning rate \(1\!\times\!10^{-4}\), weight decay \(0.01\)) for 3 epochs. We use a per-device batch size of \(1\) with gradient accumulation over \(8\) steps (effective batch size \(8\)) and \(100\) warmup steps followed by cosine decay. All training is performed in BF16 mixed precision to balance computational efficiency and numerical stability. The baseline method (ControlNet) follows its original training protocols and is tuned within the same hyperparameter budget. For MedicalSys, we split the approximately 3,000 query--response pairs at each permission level into training, validation, and test sets with an 8:1:1 ratio. The split is performed at the medical-record level, ensuring that all records used for testing are completely unseen during training and validation.

\paragraph{Inference.}
At inference time, we load the trained \(\Delta\Theta\) and apply \texttt{Intervene} at layer \(\ell\) during a single forward pass of the frozen backbone (Algorithm~\ref{alg:permit}). All methods generate autoregressively with deterministic decoding (greedy, \texttt{do\_sample=False}, temperature \(=0\)) to remove sampling variance from the comparison.

\paragraph{Hardware and reproducibility.}
All experiments are conducted on a single NVIDIA A100 (80\,GB) GPU, with each training run completing in approximately two hours. We fix random seeds for model initialization and CUDA operations to ensure reproducibility across all baselines and ablation studies. 

\subsection{Metrics}
We use string matching to compare predicted factual fields with the ground truth. For each factual field, tokens in the prediction that also appear in the corresponding ground-truth field are counted as true positives, unmatched predicted tokens as false positives, and unmatched ground-truth tokens as false negatives. Precision, recall, and F1-score are then computed from these string-level counts. For leakage rate, a response is considered leaking if it contains any string-level match with a ground-truth field that is not permitted under the given permission constraint.

\subsection{Permission Control: A Case Study}
We present a qualitative case study to illustrate the practical access-control behavior of \textsc{Permit}. 
The input query asks for highly sensitive personal information, including the patient's name, gender, age, ID number, telephone number, and address. 
The associated medical record contains both identity attributes and clinical information.

As shown in Figure~\ref{fig:permit_case_study}, \textsc{Permit} exhibits a clear permission-aware disclosure pattern. 
At the lowest permission level (L1), the model returns only a generic statement and withholds all patient-specific details. 
At L2, it reveals limited non-identifying demographic and symptom information (e.g., age, gender, and clinical condition), while still suppressing direct identifiers. 
At L3, the model further discloses partially sensitive information such as the patient's name, age, and gender, but continues to withhold highly sensitive identifiers including the ID number, telephone number, and detailed address. 
At the highest level (L4), the model provides the most complete response, disclosing both identity and medical information.

This example demonstrates that \textsc{Permit} can enforce graded access control in generation: lower permission levels preserve privacy by restricting sensitive attributes, whereas higher permission levels enable broader disclosure when authorized.

\section{Broader Impact Statement}
\label{app:broader_impact}


\textsc{Permit} has significant potential to advance the safe deployment of LLMs in privacy-sensitive sectors such as healthcare, finance, and law by regulating how a model \emph{uses} accessible information rather than merely \emph{whether} information is reachable. Its support for multi-role, hierarchical permissions with a tunable security--utility trade-off mirrors the structure of real-world organizational access policies far more faithfully than methods enforcing a single global safety boundary. The lightweight design lowers the compute and energy cost of adapting LLMs to new policies and broadens access for groups with limited resources. As an inference-time intervention complementary to existing data-level techniques, \textsc{Permit} also fits naturally into defense-in-depth deployments.

However, these benefits come with risks. The same representation-level steering could be repurposed for censorship or biased moderation, and any biased policy specified by a deployer will be enforced just as faithfully as a legitimate one. Strong leakage reduction also does not amount to a formal guarantee: adversarial prompts or out-of-distribution inputs may still elicit restricted content, and overly aggressive interventions may cause over-refusal of legitimate queries. We mitigate these concerns by recommending that \textsc{Permit} be deployed with transparent, auditable, and bias-reviewed permission policies, treated as one layer of a defense-in-depth pipeline rather than a standalone safeguard, and paired with human oversight for high-stakes decisions. We encourage future work on adversarial robustness, certified bounds, and transparent policy specification.

\begin{figure}[H]
\centering
\footnotesize
\setlength{\tabcolsep}{4pt}
\renewcommand{\arraystretch}{1.15}

\begin{tabularx}{\textwidth}{>{\raggedright\arraybackslash}p{0.17\textwidth} X}
\toprule



\textbf{User Query:} &
What is the patient's personal information, including \sens{name}, \sens{gender}, \sens{age}, \sens{id\_number}, \sens{tel}, and \sens{address}?  \\

\midrule

\textbf{Medical Record:} &
Case ID: 3202 \\
&
Name: \sens{Chang Xiang} \\
&
Gender: \sens{female} \\
&
Age: \sens{35} \\
&
ID number: \sens{330329196210214406} \\
&
Tel: \sens{15243247589} \\
&
Address: \sens{Block S, Shenhetian Road, Ling City, Shaanxi Province} \\
&
Hospital: First People's Hospital of Hangzhou City \\
&
Visiting department: Oncology Surgery \\
&
Date of consultation: March 18, 2024 \\
&
Chief complaint: The patient complained of persistent pain in the right lower abdomen in the past month, accompanied by symptoms such as loss of appetite and weight loss. \\
&
Current medical history: The patient began to experience right lower abdominal discomfort without obvious inducement nearly a month ago, which gradually worsened, accompanied by intermittent dull pain. At the same time, the patient reported a significant decrease in appetite and a weight loss of about 5 kilograms in the past month. The patient had no other symptoms such as fever, nausea and vomiting. \\
&
Past history: The patient denied a history of chronic diseases such as hypertension and diabetes, a history of surgery and trauma, and a history of drug allergies. There is no family history of genetic disease. \\
&
Physical examination: Body temperature: 36.8$^\circ$C; blood pressure: 120/80 mmHg; heart rate: 78 beats/min; respiratory rate: 18 beats/min; abdominal examination showed mild tenderness in the right lower abdomen. \\
&
Auxiliary inspection: Blood routine: normal white blood cell count; abdominal B-ultrasound: suspected space-occupying lesion in the right lower abdomen; tumor marker testing: CA19-9 slightly increased. \\
&
Preliminary diagnosis: a space-occupying lesion in the right lower abdomen, considered as a possible tumor. \\
&
Handling opinions: hospitalization for further observation and treatment; further examinations such as CT or MRI; oncology surgery consultation; rest well, avoid strenuous exercise, and maintain good eating habits. \\



\toprule
\textbf{Permission Level} & \textbf{Response} \\
\midrule

\textbf{L1} &
The patient's visit involved a health-related process. No specific personal details are disclosed. \\

\midrule

\textbf{L2} &
A \allowinfo{35-year-old female} with persistent pain and weight loss. No identifying details are disclosed. \\

\midrule

\textbf{L3} &
The patient is \allowinfo{Chang Xiang}, a \allowinfo{35-year-old female}. ID number, telephone number, and address are withheld.
She was seen at the Oncology Surgery department and complained of right lower abdominal pain. She has lost weight and appetite. She was advised to rest and follow up. \\

\midrule

\textbf{L4} &
The patient, \allowinfo{Chang Xiang}, is a \allowinfo{35-year-old female} from \allowinfo{Shaanxi province, residing at Block S, Shenhetian Road, Ling city}. Her contact number is \allowinfo{15243247589}. She visited the First People's Hospital of Hangzhou City on March 18th, 2024. She was diagnosed with a space-occupying lesion in the lower abdomen and was advised to undergo further tests, including CT or MRI scans. She was also instructed to rest, eat well, and avoid strenuous activities. \\

\bottomrule
\end{tabularx}

\caption{
We present a qualitative example of permission-controlled generation on a medical query.
Sensitive attributes in the input are marked in \textcolor{red!80!black}{red}, while model outputs that are permitted to be disclosed under each permission level are highlighted in \colorbox{green!20}{green}.
Given the same user query and medical record, the model exhibits progressive disclosure across permission levels: L1 returns only a generic non-sensitive summary; L2 reveals limited demographic and symptom information; L3 further discloses basic identity information while withholding direct identifiers; and L4 returns the most complete response. The medical record is from MedicalSys~\citep{DBLP:journals/corr/abs-2504-09593}.
}
\label{fig:permit_case_study}
\end{figure}

Overall, we are confident that the benefits of fine-grained, permission-aware generation substantially outweigh the risks, and that representation-level access control is a constructive step toward responsibly unlocking LLMs for privacy-sensitive applications.



\end{document}